\documentclass[a4paper,fleqn,usenatbib]{mnras}
\usepackage{amsmath}
\usepackage{txfonts}
\usepackage{caption}
\usepackage{comment}
\usepackage{subcaption}
\usepackage[T1]{fontenc}
\usepackage{ae,aecompl}
\hypersetup{colorlinks, citecolor=green, filecolor=black, linkcolor=blue, urlcolor=blue }

\usepackage{graphicx}
\usepackage{amssymb}
\usepackage{dblfloatfix}
\usepackage{breqn}
\usepackage{hyperref}
\hypersetup{
    colorlinks=true,
    linkcolor=blue,
    filecolor=magenta,      
    urlcolor=cyan,
}

\title[The UV/X-ray Relation in Ark 120]{Origins of the UV/X-ray Relation in Arakelian 120}

\author[Mahmoud et al.]{
R. D. Mahmoud\thanks{E-mail: mahmoud.raad@yahoo.co.uk}$^1$,
C. Done $^1$, D. Porquet$^2$ and A. Lobban$^3$
\\
$^1$ Centre for Extragalactic Astronomy, Department of Physics, University of Durham, South Road, Durham DH1 3LE, UK\\
$^2$ Aix-Marseille Univ., CNRS, CNES, LAM, Marseille, France \\
$^3$ Astrophysics Group, School of Physical and Geographical Sciences, Keele University, Keele, Staffordshire ST5 5BG, UK
}
\date{Accepted XXX. Received YYY; in original form ZZZ}

\pubyear{2022}

\hypersetup{draft}
\begin{document}
\label{firstpage}
\pagerange{\pageref{firstpage}--\pageref{lastpage}}
\maketitle

\begin{abstract}
We explore the accretion geometry in Arakelian~120 using intensive UV and X-ray monitoring from \textit{Swift}. The hard X-rays ($1-10$~keV) show large amplitude, fast (few-day) variability, so we expect reverberation from the disc to produce UV variability from the varying hard X-ray illumination. We model the spectral energy distribution including an outer standard disc (optical), an intermediate warm Comptonisation region (UV and soft X-ray) and a hot corona (hard X-rays). Unlike the lower Eddington fraction AGN (NGC~4151 and NGC~5548 at $L/L_{Edd}\sim 0.02$ and $0.03$ respectively), the SED of Akn~120 ($L\sim 0.05L_{Edd}$) is dominated by the UV, restricting the impact of reverberating hard X-rays by energetics alone. Illumination from a hard X-ray corona with height $\sim10~R_g$ produces minimal UV variability. Increasing the coronal scale height to $\sim 100~R_g$ improves the match to the observed amplitude of UV variability as the disc subtends a larger solid angle, but results in too much fast variability to match the UV data. The soft X-rays (connected to the UV in the warm Comptonisation model) are more variable than the hard, but again contain too much fast variability to match the observed smoother variability seen in the UV. Results on lower Eddington fraction AGN have emphasised the contribution from reverberation from larger scales (the broad line region), but reverberation induces lags on similar timescales to the smoothing, producing a larger delay than is compatible with the data. We conclude that the majority of the UV variability is therefore intrinsic, connected to mass accretion rate fluctuations in the warm Comptonisation region.
\end{abstract}




\section{Introduction}
\label{sec:INTRODUCTION}

The emission from Active Galaxies Nuclei (AGN) and Quasars is typically variable even in radio-quiet (non-jet dominated) objects, with the fastest timescales seen at X-ray energies. Light travel time sets a minimum size scale, and it was this, coupled to the large luminosities, which led to the first identification of the central object with an accreting supermassive black hole. 

The origin of the X-ray emission in AGN is not well understood. Standard disc models predict a maximum disc temperature which is too low to produce much X-ray flux, peaking instead in the (unobservable) far ultraviolet (FUV)/extreme ultraviolet (EUV) for most masses, spins and accretion rates. Nonetheless, stellar mass accreting black holes in our Galaxy also produce X-ray emission at energies above their disc peak, where it is generally assumed to be produced by hot, optically thin plasma which is either above an optically thick disc (sandwich corona), or is on the spin axis of the black hole (lamppost) or replaces the disc in the inner regions (truncated disc/hot inner flow). Additionally in AGN (and also in some of the more complex spectra seen from stellar mass black holes) there is another spectral component which spans between the disc and hot corona. In AGN this is typically seen as a `soft X-ray excess'. a rise in the spectrum above the low energy extrapolation of the power law spectrum seen in $2-10$~keV. This can be well fit by warm, optically thick Comptonisation (\citealt{P04,GD04}), or alternatively by extremely strong relativistically smeared reflection of the power law from the disc (\citealt{C06}). Some fraction of the soft X-rays reverberate, so must be from reprocessing as they follow the variability of the hard X-ray power law but with a lag (e.g. \citealt{K13, U14}). However, the majority of the soft X-ray excess is now thought to be a true additional component (\citealt{MBR11}, 2013; \citealt{M14,BRP16,POP18,Porquet2021}).

Going further down in energy, the spectrum which emerges from interstellar absorption in the UV is often very blue, but not as blue as expected from the outer radii of a standard accretion disc. Pure disc models have $F_\nu\propto \nu^{1/3}$, while average near UV spectral slopes (from SDSS spectra for $0.7<z<1.1$) are $F_\nu\propto \nu^{-1/3}$ (\citealt{XLH16, DWB07}). Some part of this may be due reddening from dust local to the AGN environment, but these redder slopes are clearly intrinsic in some AGN, and this UV downturn can connect to the warm Comptonisation models for the soft X-ray excess (e.g. \citealt{MBR11}; \citealt{D12}; \citealt{J12}; \citealt{M14}; \citealt{POP18}; \citealt{KD18}).  

These spectral observations can be tied together to inform a potential structure for the accretion flow, where the emission thermalises as in a standard disc only in the outer regions. Inwards of some radius, $R_{warm}$, thermalisation is incomplete, so the disc emits as warm Comptonisation rather than blackbody flux. Then at $R=R_{hot}$, the disc disappears completely, being replaced by a hot inner flow (\citealt{D12}; \citealt{KD18}). 

This three-component approach can also explain the observed trends in spectral shape with mass accretion rate (scaled to the Eddington luminosity, hereafter $L/L_{Edd}$) with a single additional assumption that the hot corona always has $L/L_{Edd}=0.02$, which is the maximal luminosity for an advection dominated accretion flow (\citealt{KD18}). This sets the radius, $R_{hot}$, within which all the gravitational energy of the flow needs to be dissipated in heating the hot plasma. For AGN with $L_{bol}/L_{Edd}<0.02$, the entire inner disc is replaced by a hot flow. The lack of a strong UV emitting inner disc means that the UV flux illuminating the broad line region (BLR) is much reduced, so the broad permitted line signature of AGN activity is likewise much less evident. This can explain the abrupt transition of `changing state' AGN, where the UV drops abruptly below $L/L_{Edd}=0.02$, in a very similar manner to the abrupt drop in disc flux in the stellar mass black hole binaries as they dim below this luminosity (\citealt{ND18}; \citealt{RAE19}). Above $L_{bol}/L_{Edd}\sim 0.02$ there is some residual UV emission from the outer disc/warm Compton disc, but the majority of the accretion energy is dissipated in the inner hot flow. It is only for $L_{bol}/L_{Edd}> 0.1-0.2$ that the hot flow is less than $10\%$ of the total power. This decreasing ratio between the hot X-ray emission and the optically thick disc, ($L_x/L_{UV}$), with increasing $L/L_{Edd}$ matches the quasar results of \cite{LR17}, and also implies that there are more seed photons cooling the hot plasma, steepening the hard X-ray spectral index with increasing $L/L_{Edd}$ as observed.

This is a specific geometry, which is testable using the intensive reverberation campaigns, where optical, UV and X-rays are monitored simultaneously with the {\it Swift} satellite. These large campaigns show differential lags across the UV and optical, with the FUV responding first, followed by longer wavelengths with progressively longer lags (e.g. \citealt{E15}, 2017, 2019). However, these lag timescales are typically larger than expected from illumination of the outer disc by a central X-ray source, irrespective of whether the disc is somewhat truncated on its inner edge and/or covered by a warm Compton region as expected in the radially stratified models of the accretion flow described above (\citealt{GD17, MD18a}). The data require much larger size scales for the reprocessor if light travel time sets the lags. There is some convergence in the literature that there is an additional diffuse emission component, either from the broad line region itself (\citealt{KG01}, 2019; \citealt{LGK18}) or from a wind inwards of the broad line region (\citealt{DFP19, Kara2021}). This is seen directly as an increased lag around the Balmer continuum (\citealt{Cackett2018, Edelson2019, Cackett2022}). The longer lags could also indicate the X-ray source is not central, but is instead at a much larger distance from the accretion disc, along the black hole spin axis (\citealt{KammounPapadakisDovciak2019},~2021). However, a more fundamental problem with both these solutions is that the UV is poorly correlated with the X-rays which are meant to be driving the whole variability (e.g. \citealt{Edelson2019}). 

X-ray variability can be complicated. There can be spectral pivoting (a steeper power law when the spectrum is brighter) so that the bolometric flux varies differently to that in the observed (generally $2-10$~keV) X-ray bandpass. There can also be absorption variability along the line of sight, which likewise means that the observed X-ray flux does not track the intrinsic variability seen by the disc. Variable X-ray absorption is particularly an issue in NGC~5548, the first AGN used for these campaigns, as it was unusually obscured during the monitoring (\citealt{MKK16}; \citealt{DFP19}). Similar issues are now clearly present in Markarian 817 (\citealt{Kara2021}). 

Hard X-ray monitoring circumvents both these uncertainties, as it  much less affected by the absorption variability. However, NGC~4151 is the only AGN which is bright enough for Swift BAT monitoring above $10$~keV, and this showed clearly that the disc was strongly truncated, with no optically thick material responding within a few hundred $R_g$ (\citealt{MD20}, hereafter MD20). This was independently confirmed by studies of the iron line profile which is marginally resolved in Chandra HETGS (\citealt{Miller2018}), indicating an origin at inner broad line region scales, as also supported by newer iron line reverberation studies (\citealt{ZMC19}). This is consistent with the models described above, as NGC~4151 was below $L=0.02~L_{Edd}$ during the campaign, so should not have an inner UV bright disc. Nonetheless, there \textit{is} a potential reprocessing signal in the UV in these data. It is instead consistent with material in the inner broad line region. A reprocessed continuum component from the densest clouds in the BLR is expected theoretically (\citealt{KG01}, 2019; \citealt{LGK18}). All sketches of the BLR have it subtending a much larger solid angle to the central source than the flat disc, and its reprocessed signal fits the timescales seen in the reverberation data (\citealt{KG19}).

There is another way to circumvent the uncertainties in going from observed X-ray $2-10$~keV flux to bolometric, and that is to use one of the `bare' AGN (\citealt{PAT11, WAL13}) which: a) show no absorption in their soft X-ray spectra, so that absorption variability is not an issue, and b) provide good enough statistics in the X-ray bandpass up to $10$~keV that one can model the effects of spectral pivoting. This selects Ark~120, which spans $L/L_{Edd}\sim 0.04-0.06$, so according to the models above should have some outer disc/warm Compton component to explore in reverberation (as well as the BLR/inner wind). We use the recent {\it Swift} UV and X-ray monitoring, combined with XMM-Newton/NuSTAR data to determine the spectral components and their variability, and use this to determine limits on the accretion geometry at this brighter $L/L_{Edd}$.

\begin{figure*}
	\includegraphics[width=15cm]{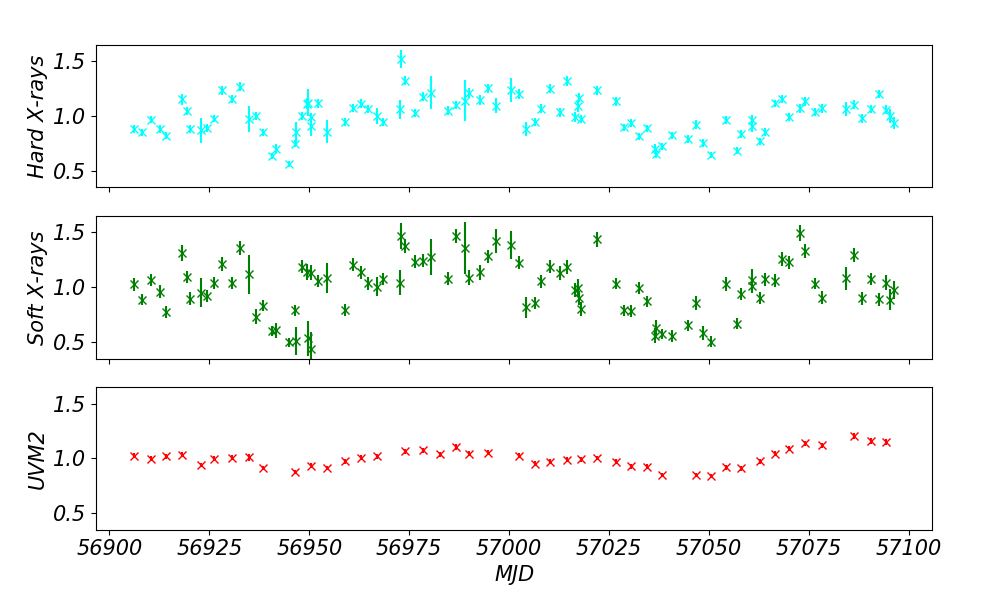}
	\caption{The raw, mean normalised light curves for this campaign in the bandpasses of the hard X-ray ($1-10$~keV, uppermost panel), soft X-ray ($0.3-1$~keV, middle panel) and UVM2 (lower panel).}
	\label{fig:lightcurves_alone}
\end{figure*}

\section{Data}
We need both good quality spectra and good quality lightcurves to build a spectral-timing model where the reprocessed variability in optical and UV can be predicted from the observed X-ray flux. We choose the combination of XMM-NuSTAR  spectra from 2013-02-18 and 2014-03-22, and Swift XRT-UVOT monitoring data covering a $\sim 6$-month period, from 2014-09-04 to 2015-03-15 (\citealt{PRM18}, hereafter P18; \citealt{LPR18}, hereafter L18, see also \citealt{Gliozzi2017}, \citealt{BLA17}). While the XMM-NuSTAR snapshots do not overlap in time with the Swift monitoring, the lowest and highest points on the X-ray lightcurves in the SWIFT monitoring match the X-ray flux from the 2013 (low) and 2014 (high) XMM-NuSTAR spectra (P18, L18) so we use these to define the spectral range.

\begin{figure}
\centering
    \includegraphics[width=0.9\hsize]{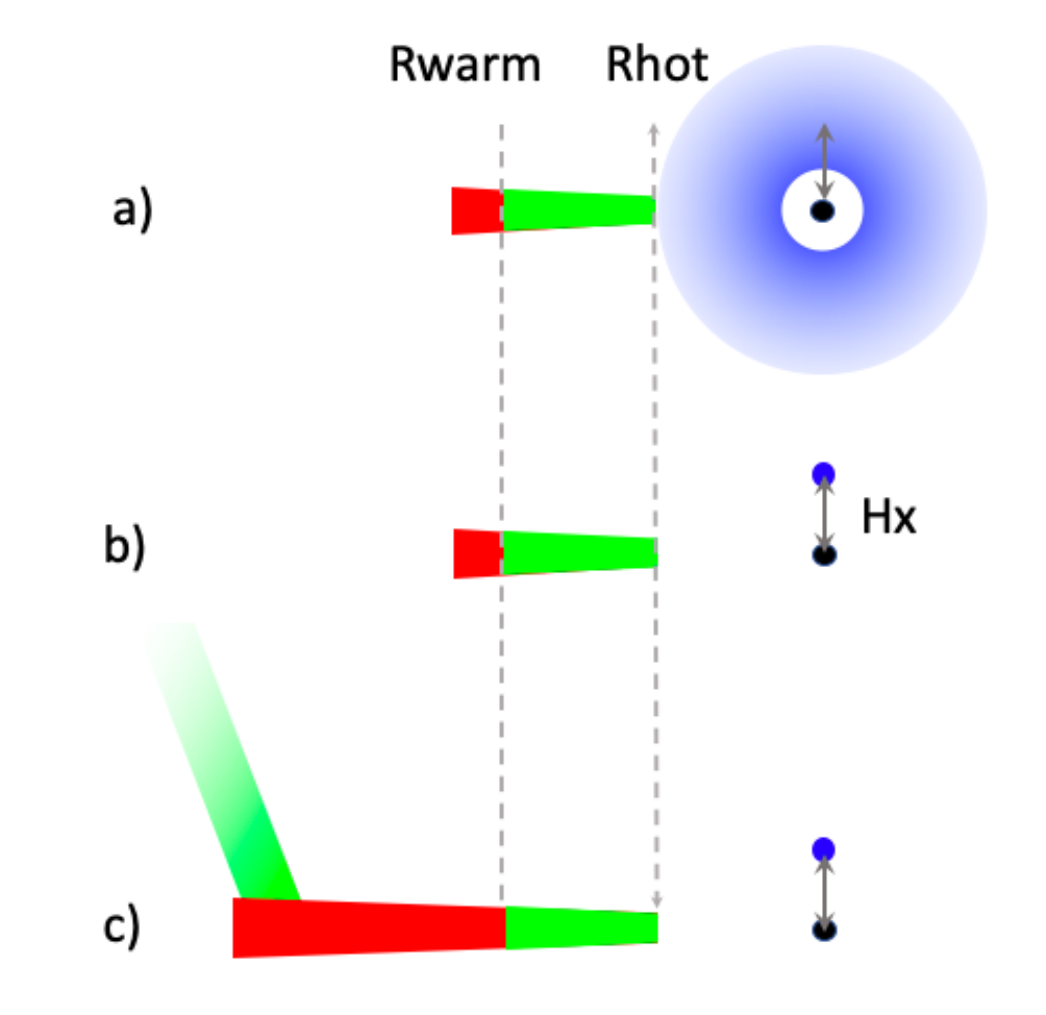}
    \caption{The geometries assumed in this paper a) the intrinsic components of the {\tt agnsed} model. b) the extended hard X-ray Comptonising region is assumed to be a point source at height $H_x$ above the black hole on the spin axis for ease of calculation of the reprocessed emission. c) Includes the additional reprocessor connected to the inner BLR and/or its wind.} 
    \label{fig:geom}
\end{figure}

\subsection{Swift light curves}
There are a total of $86$ SWIFT observations in the 2014 campaign (ObsID: 00091909XXX), each separated by $\sim2$~days, with typical duration of $\sim1$~ks. The U and UVM2 filters were alternated, so each UV filter point has typical separation of $\sim4$~days (L18). We choose to focus on the shortest wavelength band, UVM2, as this is least affected by host galaxy and BLR contamination. 

Fig~\ref{fig:lightcurves_alone} shows the mean-normalised X-ray and UVM2 lightcurves. The XRT data is split into two energy bands, the hard X-rays from $1-10$~keV (hereafter HX) and soft X-rays from $0.3-1$~keV (hereafter SX), from L18. Each dataset has been normalised to its mean count rate, showing clearly that there is more variability on these timescales in HX and SX than in UVM2, but that the lightcurves are fairly well correlated by eye on these timescales. The error-subtracted hard X-ray variance is $0.174$, while the soft X-ray variance is larger, at $0.250$. This already presents a challenge for models where all variability is reprocessed hard X-ray flux, although it could perhaps be produced by temperature variability of the warm Comptonisation amplifying the response over a small soft X-ray band. 

\subsection{ARK 120 spectral energy distribution}
\label{SED fitting}
The XMM-NuSTAR observations from P18, which includes XMM-OM UV data, gives the full spectral energy distribution (SED) at two epochs. We fit the spectra from these two epochs jointly in {\tt{xspec}}. To do this, we follow MD20 by applying the {\tt{agnsed}} model of \cite{KD18}, which consists of an outer standard disc (red lines), a warm Comptonised disc (green lines) and (blue) hard Compton components. In addition to this, we include neutral reflection using {\tt{pexmon}}, relativistically smoothed with {\tt{rdblur}} representing mildly blurred disc reflection (N16, P18, P19). The more detailed spectral study in P18 uses the more sophisticated {\tt{relxill}} reflection models, but their best fit is only mildly ionised so the simpler neutral reflection model is appropriate for our purpose which is simply to extract the underlying continuum. We also include a free narrow Gaussian component to model the ionised emission line seen at $6.95$~keV, and one to match the slightly broadened $6.4$~keV line which is likely from the BLR/inner wind (P18). The total model is absorbed by the interstellar medium using {\tt{tbfeo}} (\citealt{WAM00}) for gas and  {\tt{redden}} (\citealt{CCM89}) for dust so the total model is {\tt{constant * tbfeo * redden * (rdblur * pexmon + agnsed + zgauss + zgauss)}}. The assumed geometry for the {\tt agnsed} model is shown in Fig.~\ref{fig:geom}a. It is noteworthy that this model assumes a real truncation of the optically thick disc at $R_{hot}$, so predicts an iron line from disc reflection which is not highly relativistically smeared. This is consistent with more detailed reflection spectral fits to the data, as discussed in P18.

The {\tt agnsed} model also includes the reprocessed hard X-ray emission. \cite{GD17} show explicitly that an extended hard X-ray source with emissivity like that expected from a thin disc gives an illumination pattern that is extremely similar to that from a much simpler point source at height $H_x=10~R_g$ above the spin axis (Fig.~\ref{fig:geom}b).

\begin{figure*}
    \includegraphics[width=0.7\hsize]{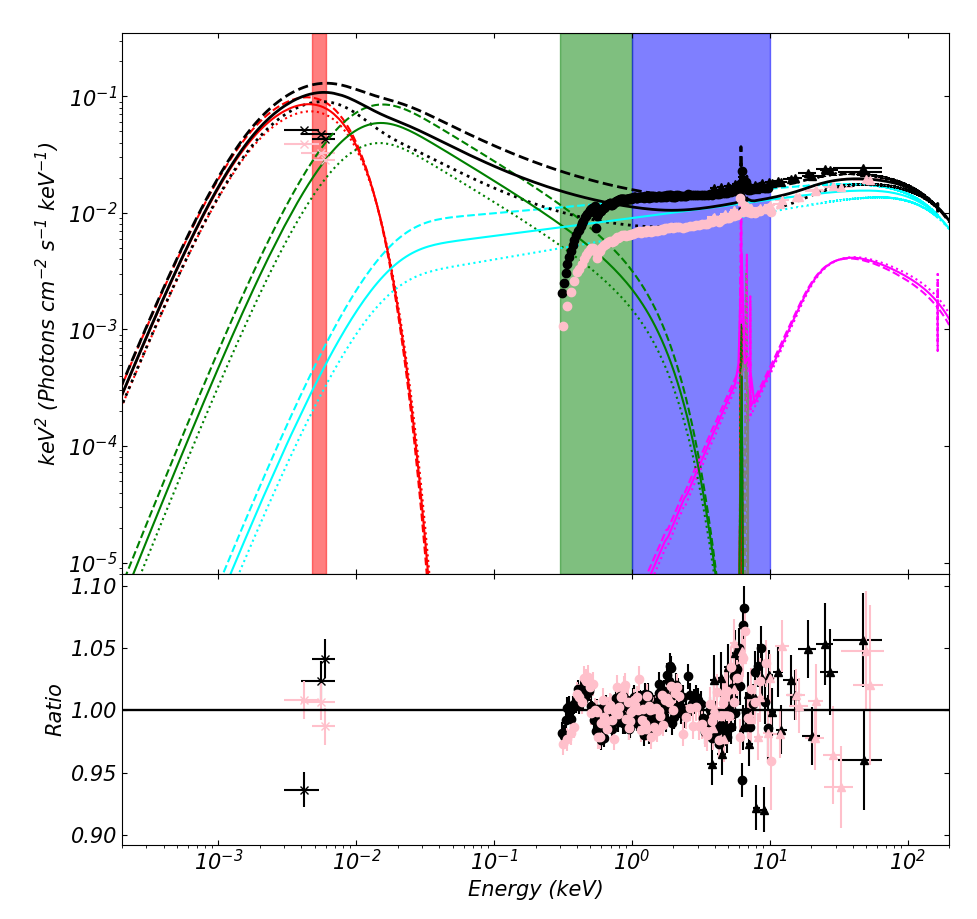} 
    \caption{Top panel: broad-band spectrum and SED for the XMM-NuSTAR observations of P18. The red vertical line indicates the UVM2 band, while the green and blue show the soft and hard X-ray bands. The black points show the observed (i.e. absorbed) high-luminosity (2014) epoch. The pale pink points are the low-luminosity (2013) epoch (absorbed). The (unabsorbed) best fit model to the 2014 observations is shown as the black dashed line with components in dashed red (SS disk), dashed green (soft Compton), and dashed turquoise (hard Compton). The (unabsorbed) best fit model to the 2013 observations is shown as the dotted black line with components in dotted red (SS disk), dotted green (soft Compton) and dotted turquoise (hard Compton). The average spectrum between the two epochs is shown as the solid black line, with solid components (colours correspond to the same as individual epochs, above). We take this spectrum to be the assumed mean spectrum during the \textit{Swift} monitoring campaign. Relativistic reflection and the free Gaussian lines are also shown in pink and grey respectively, where the line styles denote each epoch and the mean as above. Bottom panel: Residual ratios of the high- and low-luminosity epoch data to their best-fit spectral models (when absorption is applied).}
    \label{fig:SEDs_with_data}
\end{figure*}

Fig.~\ref{fig:SEDs_with_data} shows the observed data and unabsorbed model fit lines from the high epoch (black circles, dashed lines) and low epoch (pink circles, dotted lines). These are well fit, as shown by the lower residual panel, and give a total Eddington fraction of $L/L_{Edd} = 6.3\%$ and $3.8\%$ for the higher and lower luminosity epochs respectively for this black hole mass of $1.5\times 10^8M_\odot$, so both mass and mass accretion rate are slightly higher than in NGC~5548, and the SED has higher UV to X-ray luminosity showing there is some intrinsic inner disc emission
(see also \citealt{KD18}). 

To infer the lightcurve contributions from each component, we would like the mean SED during the monitoring campaign period. Fortunately, the fluxes in each of the high and low epochs match the high and low flux limits during the monitoring campaign period from 2014 to 2015. We can therefore take the mean SED during the monitoring campaign period to be the spectrum bisecting the higher and lower luminosity epochs. This intermediate spectrum is shown as the solid lines in Fig.~\ref{fig:SEDs_with_data}.

\begin{table}
    \centering
    \hspace{-2cm}
	\begin{tabular}{l|ll|ll}
				            & $h_x=10$ case             &               & $h_x=100$ case        &       \\	
		\hline
		Observation			& 2013      				&2014           & 2013                  & 2014  \\
		                    & (Low)                     &(High)         & (Low)                 & (High)\\
		\hline
		\vspace{+3pt}
		$M_{BH}$		    & $1.5\times 10^8$ (F)      & ==            & ==                    & ==    \\
		\vspace{+3pt}
		$D$  			    & $143.5$ (F)	            & ==            & ==                    & ==    \\
		\vspace{+3pt}
		$a^*$               & $0$ (F)                   & ==            & ==                    & ==    \\
		\vspace{+3pt}
		cos$(i)$			& $1.$ (F)		            & ==            & ==                    & ==    \\
		\vspace{+3pt}
		log$(\dot{m})$		& $-1.42\pm0.02$	        & $-1.20\pm0.01$       & $-1.42^{+0.03}_{-0.02}$               &$-1.23\pm0.02$\\
		\vspace{+3pt}
		$kT_{e,\,hot}$ 		& $100.$ (F)	            & ==            & ==                    & ==    \\
		\vspace{+3pt}
		$kT_{e,\,warm}$ 	& $0.361^{+0.002}_{-0.001}$   				& $0.342\pm0.002$       & $0.366\pm0.003$               &$0.348\pm0.001$\\
		\vspace{+3pt}
		$\Gamma_{hot}$		& $1.80^{+0.01}_{-0.02}$            		& $1.90\pm0.01$        & $1.80^{+0.02}_{-0.03}$                &$1.89^{+0.02}_{-0.01}$ \\
		\vspace{+3pt}
	    $\Gamma_{warm}$ 	& $2.70$ (F)                & ==            & ==                    & ==    \\
		\vspace{+3pt}
		$R_{hot}$ 			& $23.0\pm0.2$	                & $23.1^{+0.3}_{-0.2}$        & $23.1\pm0.3$                & $22.8^{+0.1}_{-0.2}$\\
		\vspace{+3pt}
		$R_{warm}$ 			& $46^{+2}_{-1}$	    & $61\pm1$        & $46.4\pm0.3$    & $63\pm1$\\
		\vspace{+3pt}
		log$(r_{out})$ 		& $-1$ (F)          		& ==            & ==                    & ==    \\
		\vspace{+3pt}
		$h_{max}$ 			& $10$ (F)                  & ==            & $100$ (F)             & ==    \\
		\vspace{+3pt}
		reprocess     		& $1$ (F) 	                & ==            & ==                    & ==    \\
		\vspace{+3pt}
		redshift 			& $0.0327$ (F) 	            & ==            & ==                    & ==    \\
		\hline
	\end{tabular}
	\caption[]{Obtained parameters for {\tt{agnsed}} from fitting to the time-averaged 2013 and 2014 epoch data. ``F" denotes a fixed parameter. First and second columns denote fit using $h_x=10$, used in most of the paper and shown explicitly in Fig.~\ref{fig:SEDs_with_data}. Third and fourth columns denote fit using $h_x=100$, used primarily in Fig.~\ref{fig:reprocessing_variablehard_flatsoft_flatdisc_LCs_HX100_corrected}.}
	\label{tab:agnsed_params}
\end{table}

\section{Light curves of the components}

The spectral index of the hard X-rays clearly varies with flux, softening as the source brightens (L18), but peaking always at high energies, above $50$~keV. Thus the bolometric hard X-ray coronal flux varies by less than the $1-10$~keV flux. We reconstruct the bolometric hard X-ray variability by using the observed spectral pivoting between the 2013 and 2014 spectra, and interpolating between them, assuming that there is a one-to-one relation between $1-10$~keV flux and spectral index (as is generally seen: \citealt{LZPS20}), and assuming that the electron temperature stays constant at $100$~keV.

\begin{figure*}
	\includegraphics[width=15cm]{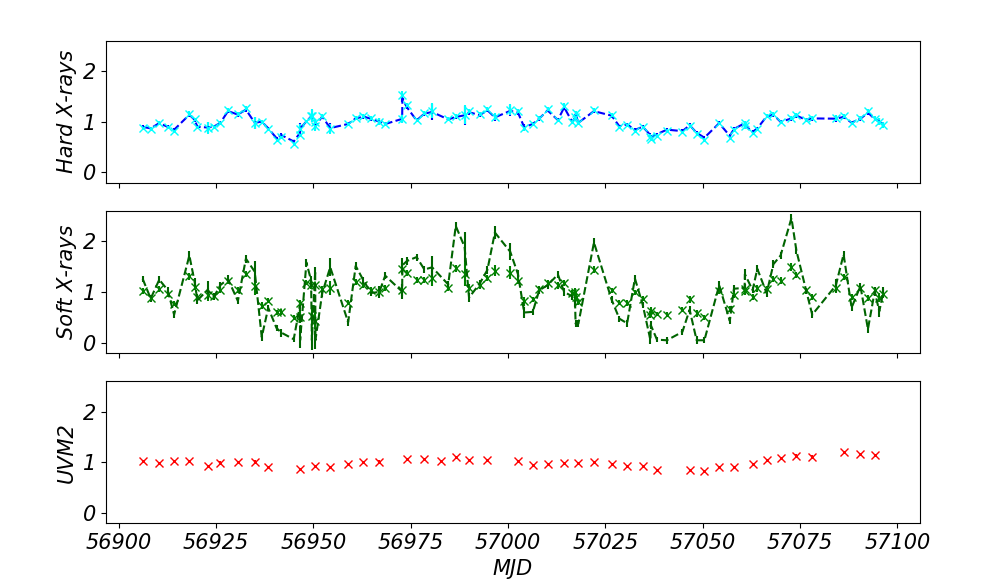}
	\caption{Crosses are the observed lightcurves in the hard X-ray, soft X-ray and UVM2 bands, as in Fig.~\ref{fig:lightcurves_alone}. The cyan dashed line shows the inferred mean-normalised variations of the hard Compton component at its bolometric peak, where the variation amplitudes are adjusted for the expected pivoting and higher mean. The green dashed line denotes the expected variations in the soft X-ray Component, assuming the non-pivoting case. This is inferred from the soft band light curve, by computing how the hard Component contribution in the soft band are expected to vary, subtracting these variations from the observed soft band curve, and mean-normalising the resultant curve. We see that the soft X-ray component is intrinsically more variable than the soft band curve itself.}
	\label{fig:lightcurves_with_component_curves}
\end{figure*}

The top panel of Fig~\ref{fig:lightcurves_with_component_curves} shows the observed mean normalised $1-10$~keV lightcurve (cyan crosses) together with the inferred mean normalised bolometric variability (dark blue dashed). The two are extremely similar, as this correction is only of order $10\%$. We also use the inferred index and normalisation to extrapolate the hot corona spectrum down, and find that the hot Compton component produces $63\%$ of the flux in the soft X-ray band. We subtract the hard variability, scaled by this proportion, from the observed $0.3-1$~keV flux so as to separate the variability of the soft X-ray excess component from the underlying variability of the hot corona. This correction has a larger effect. The light green crosses in the middle panel of Fig~\ref{fig:lightcurves_with_component_curves} show the normalised total $0.3-1$~keV lightcurve which includes both the hot corona variability and the soft excess, while the dark green shows the new estimate of the variability of the soft X-ray excess alone. This now has noticeably more variability, increasing the tension noted earlier. There must be intrinsic variability in the soft X-ray excess on these timescales, as is seen even more clearly by comparing the two soft X-ray components in the SED fitting in Section~\ref{SED fitting} (see also Fig.~$1$ in \citealt{PRM18}), where the total luminosity of the warm Comptonised component changes by a factor of $2.2$ between 2014 and 2013, while that of the hot Comptonisation changes by only a factor $1.7$. The modelled change in warm Comptonisation normalisation cannot therefore be driven by reprocessing of the hot component in a constant geometry.

We use the mean of the inferred bolometric hard X-ray variability to derive the mean hard X-ray Compton component in Fig.~\ref{fig:SEDs_with_data}. We then compute the variable reprocessing resulting from this component as it illuminates the warm Compton region and outer standard disc, assuming that these are intrinsically constant. To do this, we closely follow the timing simulation procedure of MD20, whereby the seed photon temperature at each disc annulus is modulated according to the reprocessed flux variations as
\begin{equation}
T_{seed}(r,\,t) = T_{grav}(r)\left(\frac{F_{rep}(r,\,t) + F_{grav}(r)}{F_{grav}(r)}\right)^{1/4},
\end{equation}
where $T_{grav}$ and $F_{grav}$ are the local temperature and energy flux due to gravitational dissipation, and $F_{rep}$ is the flux incident on the local annulus from the hard Compton source. The only difference in our procedure for Ark~120 is that we now include flux variations in outer thermal disc as well as the warm Compton disc, since the thermal disc component contributes significantly to the UV bandpass according to the Ark~120 SED. This is unlike NGC~4151, where the disc component was too cool to contribute to the UV. The difference in extent of the disc is as expected for the difference in $L/L_{Edd}$ between these two objects. 

The inferred bolometric X-ray lightcurve has the same sampling as the observed X-ray lightcurve, of around $2$~days, while the alternating UV filters mean that the UV sampling interval in any single band is around $4$~days. Each observation length is only $1$~ks, so the data are sampled rather than binned. Plainly this could mean that some of the fast X-ray variability is missed as the power spectrum of the $1-10$~keV lightcurve is consistent with a power law, with $PSD(f)\propto f^{-2}$ down to timescales below $1000$~s (see e.g. Fig.~$12$ in L18). We therefore use the continuous-time auto-regressive moving average (CARMA) model of \cite{KBS14} to simulate the range of variability which might be missed by the gaps in the sampled X-ray lightcurve. We then reprocess this range of possible full lightcurves from the warm Compton and outer standard disc to give the predicted UVM2 reprocessed signal in each of the models below. 

\begin{figure}
	\includegraphics[width=\columnwidth]{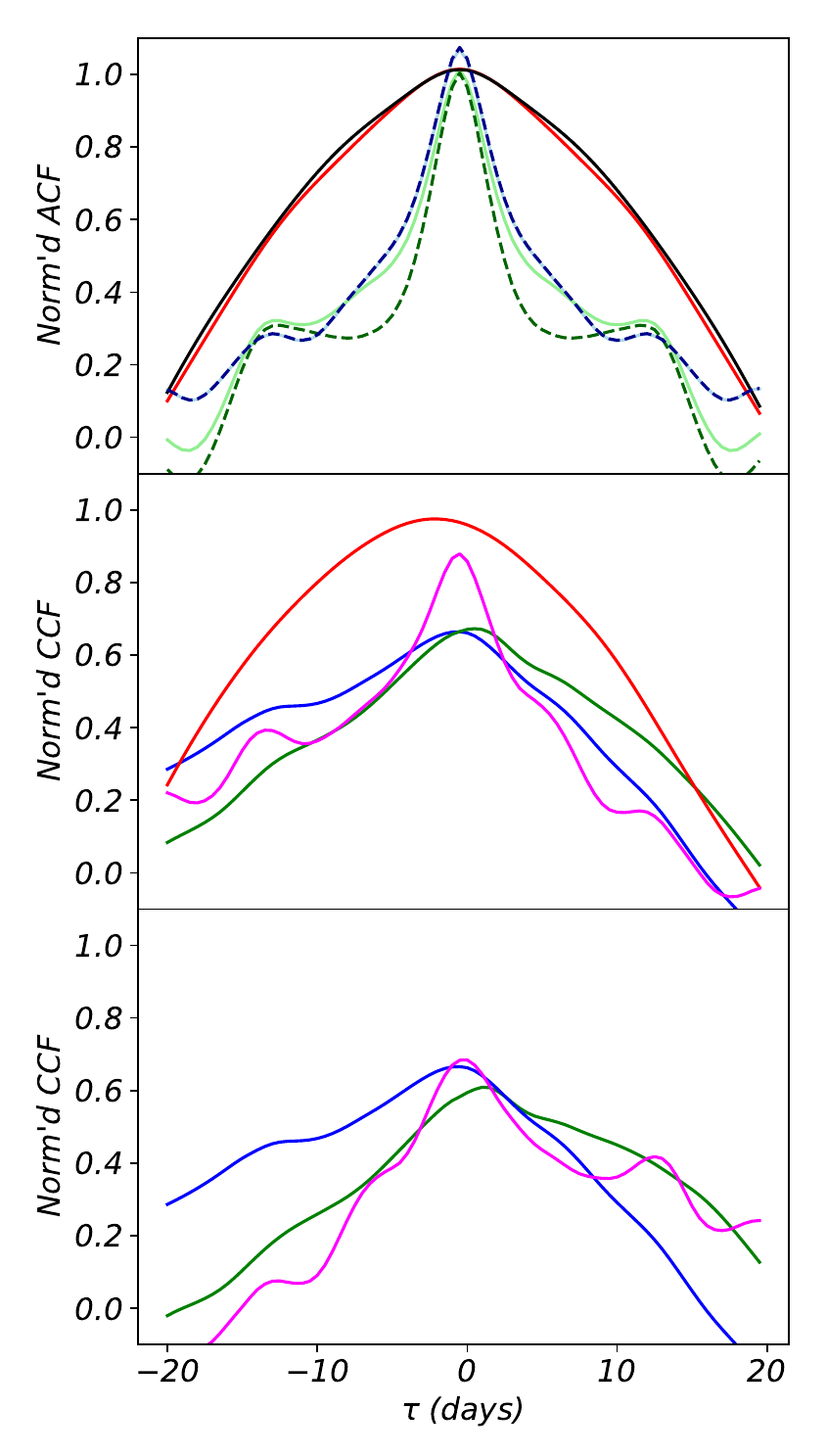}
	\caption{Top panel: Auto-correlation functions for the U band (red solid), UVM2 (black solid), soft X-ray (pale green solid), hard X-ray (pale blue solid) and derived hard Compton (blue dashed), derived soft Compton(green dashed). Middle panel: Cross-correlation functions between UVM2 count curve and: U (red solid); observed soft X-rays (light green solid); observed hard X-rays (blue). Also the cross-correlation between the observed soft X-rays and hard X-ray count curves (magenta). Bottom panel: Cross-correlation function between the UVM2 count curve and: the derived soft Compton curve (green); the derived hard Compton curve (blue). Also the CCF between the derived soft Compton and hard Compton curves (magenta).}
	\label{fig:U_UVM2_SX_HX_COMPS_CCFs}
\end{figure}

Figure~\ref{fig:U_UVM2_SX_HX_COMPS_CCFs} shows the auto-correlation functions 
of the light curves in each band. The top panel shows the auto-correlation functions (ACF) of each raw lightcurve as solid lines, with U (red) and UVM2 (black) clearly broader than the soft X-ray (light green solid) and hard X-ray (light blue solid). The ACF of the reconstructed hard Compton component (dashed dark blue) lies almost completely on top of the raw hard X-ray count rate, as it is approximately a scaled version of the observed hard X-ray variability. However, the ACF of the reconstructed soft X-ray Compton component variability (dark green dashed line) is somewhat different in shape to the soft X-ray count rate curve (solid pale green line) due to the removal of the hard X-ray variability in this band. 

The cross correlation function (CCF) of each observed lightcurve against UVM2 is shown in the middle panel. The U band (red) is extremely well correlated with UVM2 ($CCF^{max}_{UVM2, U}=0.98$), while the observed soft (green) and hard (blue) X-rays are not ($CCF^{max}_{UVM2, SX}=0.67$, $CCF^{max}_{UVM2, HX}=0.65$). The magenta line shows the cross-correlation between the observed soft X-ray and hard X-ray lightcurves. Clearly there is some correlation between these on short timescales, but much of this is due to the soft X-ray lightcurve including some of the hard X-ray variability. The magenta line in the lower panel shows the cross-correlation of the derived soft and hard X-ray Compton components, clearly showing the drop in peak at zero lag by removing the contribution of hard Compton from the soft band. The lower panel shows the CCFs of these separated soft Compton (green) and hard Compton (blue) X-ray component variability relative to UVM2. Clearly they are not well correlated. 

\section{Reprocessing of the hot Compton component} 
Here we will introduce the first spectral-timing model of this paper. In this basic case, we assume that only the hard Compton component (the corona) varies independently, and that all other variations are driven by reprocessing of this hard coronal power (see the geometry in Fig. \ref{fig:geom}a and b). This is in tension with the larger variability seen in soft X-rays than in the hard, but here we are focused instead on what this standard idea of hard X-ray reprocessing produces in the UV band.

\begin{figure}
	\includegraphics[width=\columnwidth]{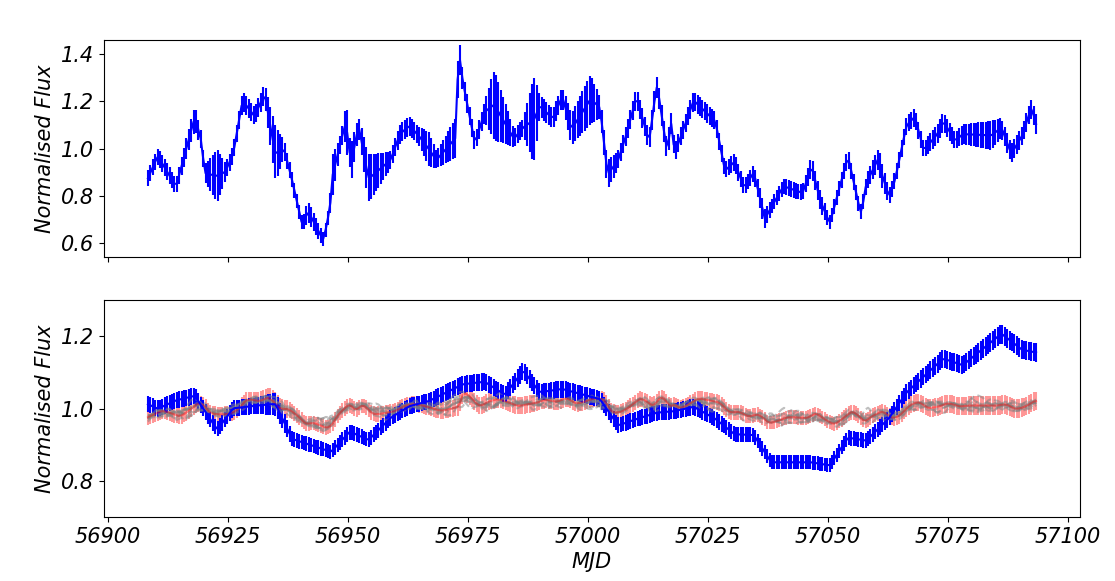}
	\includegraphics[width=\columnwidth]{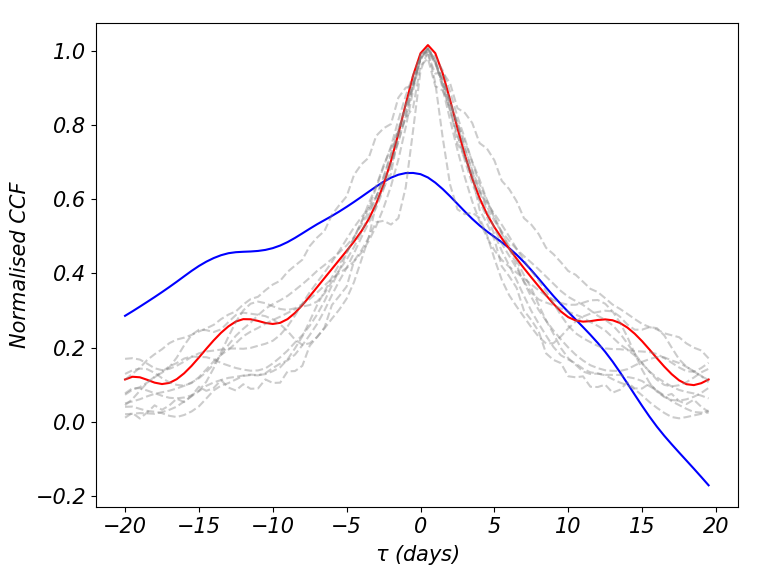}
	\caption{Results for model \textbf{HX10}. Top panel: Hard X-ray lightcurve from \textit{Swift}. Middle panel: Disc + warm Compton reprocessing of only hard illuminating flux, no intrinsic soft variability. Observed UVM2 light curve is shown in blue, with the model prediction in red assuming observed hard Compton input. Grey lines show UVM2 timing model predictions with CARMA hard X-ray lightcurve realisations as input. Bottom panel: CCFs for the observed (blue) and simulated (red) light curves, with model predictions based on CARMA hard X-ray lightcurve realisations in grey.}
	\label{fig:model_predictions_CARMA_HX}
\end{figure}

The uppermost panel of Fig.~\ref{fig:model_predictions_CARMA_HX} shows the mean normalised bolometric hot Comptonisation component lightcurve. The middle panel shows the predicted (red) and observed (blue) response in UVM2.  This was calculated assuming that the X-ray corona has mean height of $h_x=10$, as in the {\tt{agnsed}} fit, in order to calculate the illumination. This is a very poor match to the observed UVM2 lightcurve, predicting much smaller variability amplitude than is seen in the data, and an overall root-mean-square (rms) error with respect to the data of $7\times10^{-2}$. The grey lines in the middle panel show the UV curves predicted from 10 CARMA realisations of the X-ray input, displaying the range of UV curves predicted by this model which could have arisen due to under-sampling (clearly insufficient to match the observations, in blue).

The CCF of UVM2 with the hard X-rays reveals another mismatch, in that the predicted small amplitude variability is fast, and well correlated with the hard X-rays on timescales of days, as the model UV disc size is small. It predicts a peak correlation between the fluctuations in UVM2 and hard X-rays which is close to unity, and a width of around $4$~days; i.e. it looks like the ACF of the hard X-rays, while the observed UVM2 cross-correlation (blue line) is much broader and weaker. This model is clearly wrong, as the true correlation corresponds to only $CCF^{max}_{UVM2,\,HX} = 0.65$. Most of the observed UV variability is not produced by hard X-ray illumination of a disc, where the UV is produced in a region $\sim30-100~R_g$ (less than $1$~light-day) by a source of typical height $10~R_g$.

This model, with reprocessing of the hard Compton from $h_x=10$ in the disc and warm Compton zones, will be referred to as \textbf{HX10}.

\begin{figure}
	\includegraphics[width=\columnwidth]{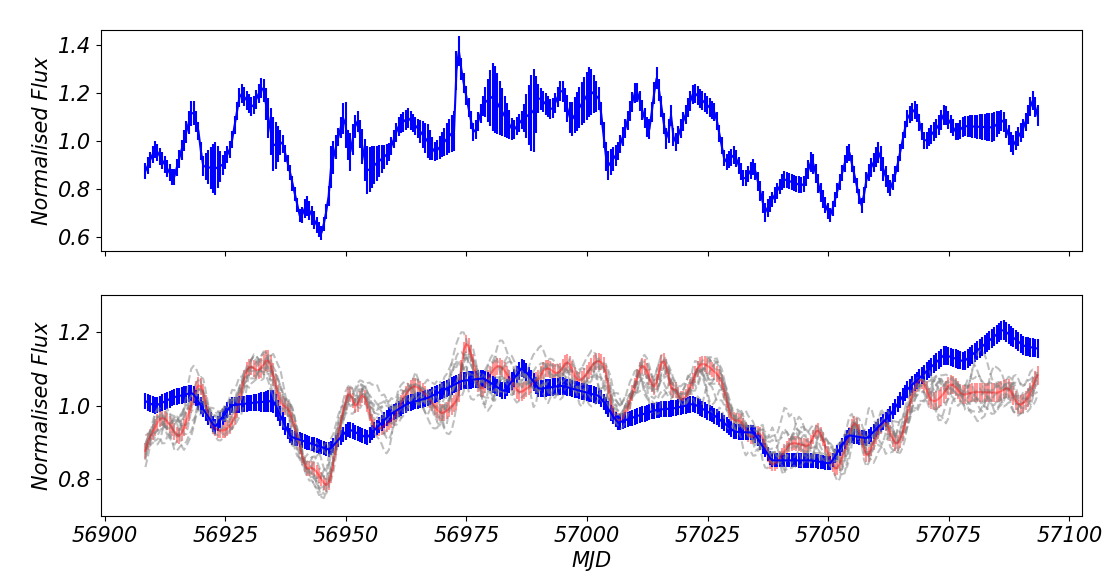}
	\includegraphics[width=\columnwidth]{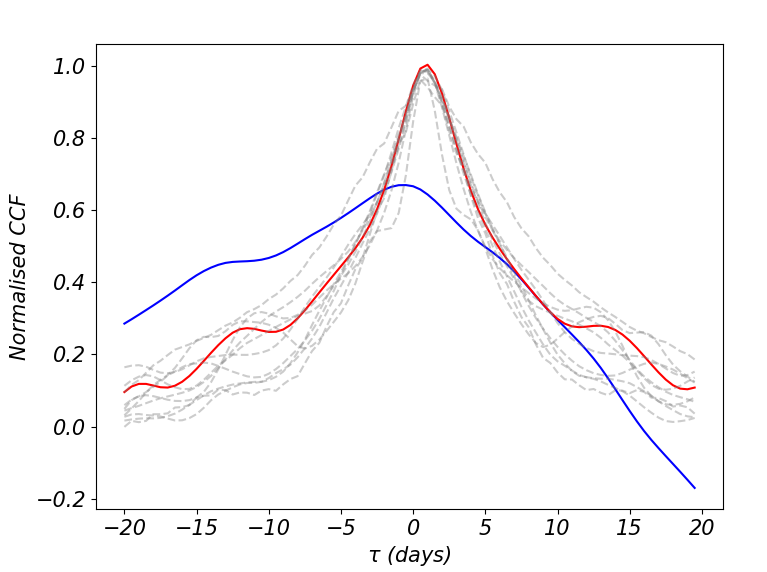}
	\caption{As in \textbf{HX10} (Fig.~\ref{fig:model_predictions_CARMA_HX}), but with the hard corona scale height set to $h_x=100$.}
	\label{fig:reprocessing_variablehard_flatsoft_flatdisc_LCs_HX100_corrected}
\end{figure}

One way to get more response in the UV from the same X-ray flux variability is to change the geometry so that the disc subtends a larger solid angle to the X-ray source. While $H_x=10~R_g$ is roughly the expected size scale of gravitational energy release, the hot corona scale height could be larger and/or the corona itself could be associated with the base of a jet at some larger scale height above the black hole (see Fig. \ref{fig:geom}b). We refit the data using an X-ray scale height of $H_x=100~R_g$. This does give more reprocessing in the UV but this is still not a large fraction of the total UV flux as the hard X-ray bolometric power is smaller than the UV. Hence the new spectral fit with {\tt{agnsed}} with height fixed to $H_x=100~R_g$ has all other parameters very similar to those derived from the original fits at $H_x=10~R_g$, and with almost identical goodness of fit ($\chi^2_{\nu}=1.14$: Table~\ref{tab:agnsed_params}).

The resultant UVM2 lightcurve is shown in Fig.~\ref{fig:reprocessing_variablehard_flatsoft_flatdisc_LCs_HX100_corrected}. This again assumes only reprocessing of the hard Compton fluctuations on the blackbody/warm Compton disc zones, but now the amplitude is comparable to that seen in the data. However, the predicted UVM2 lightcurve is not a good match to the observed UVM2 variability. The predicted lightcurve has too much fast variability compared to the observed data.

It is clear that the disc as assumed in the $H_x=100~R_g$ {\tt{agnsed}} model - where there is optically thick material at $R>23~R_g$ (though most of the response comes from $R\sim H_x=100~R_g=1$~light day) - is too close to the X-ray source for light travel time delays to smooth out the observed fast X-ray fluctuations to a level compatible with the rather smooth ($20$~light day ACF width) UV variability.

While this model initially looked promising (see also \citealt{NMW16} on NGC~3516 where they show that this model can work with sparse sampling of the X-ray and optical lightcurves), this is not a good match to this data in detail, with the rms error with respect to the observations at $7\times10^{-2}$, similar to the $h_x=10$ case. Further, the CCF shows explicitly that the predicted correlation on short timescales ($4$~days) is much stronger than seen in the real data. Thus while a large scale height X-ray source could match the UVM2 variability amplitude, it predicts too much correlated variability on short ($4$~day) timescales. This is consistent with the transfer functions shown for different scale height sources in Kammoun, Papadakis \& Dov\u{c}iak (2019; their Fig.~1), where the UV response peaks below 1~ld for NGC~5548, which translates to a few light days in the higher mass, higher mass accretion rate of Akn~120.

The clear result of this modelling is that most of the UV variability can not arise from reprocessing of the hard X-rays on the accretion disc as this predicts too much fast variability on timescales of a few days when matched to the detailed shape of the UV lightcurve rather than just the mean lags.  

\section{Reprocessing of hot Compton plus intrinsic variability of warm Compton}
\label{sec:model2}

\begin{figure}
	\includegraphics[width=\columnwidth]{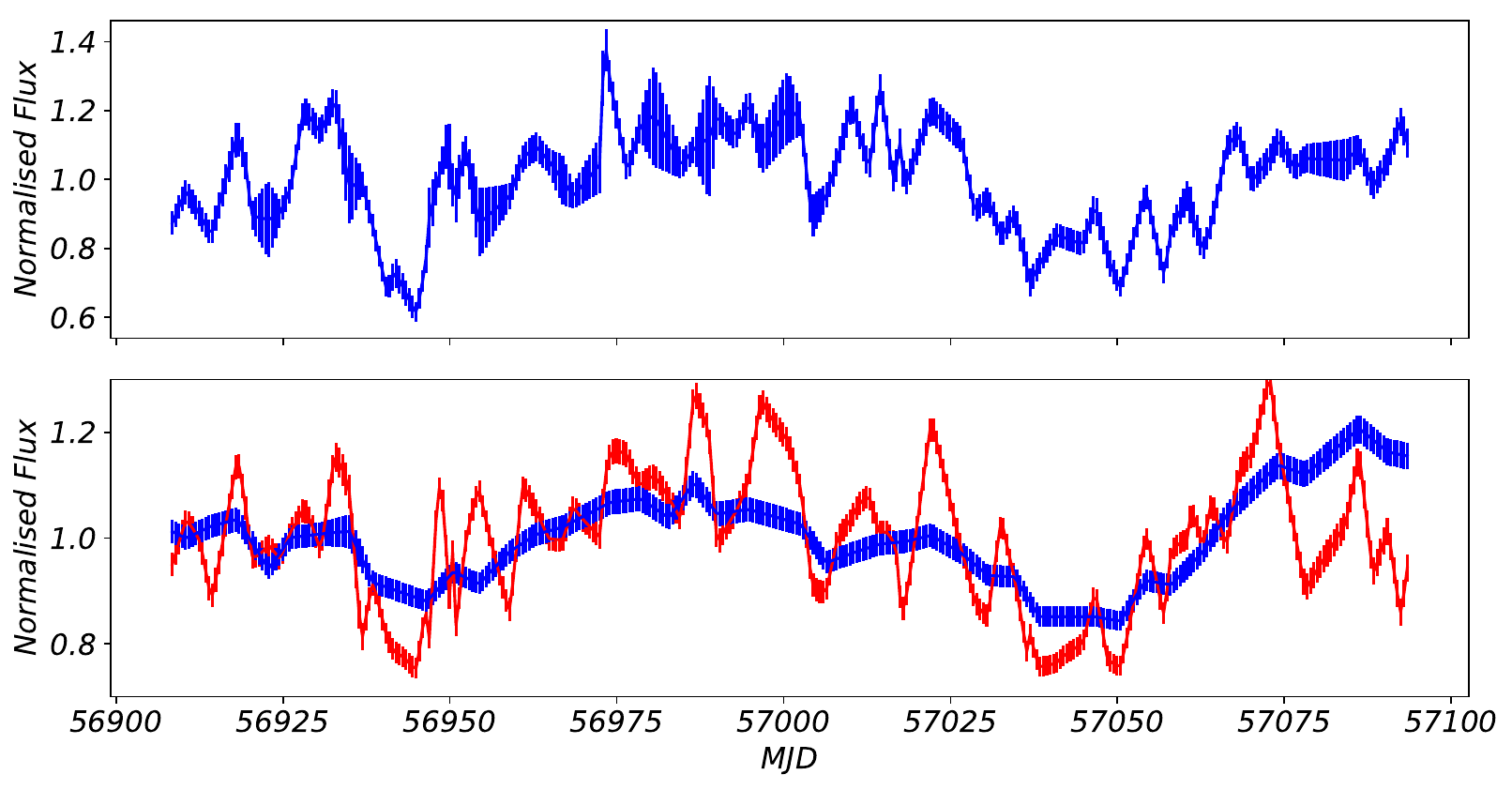}
	\includegraphics[width=\columnwidth]{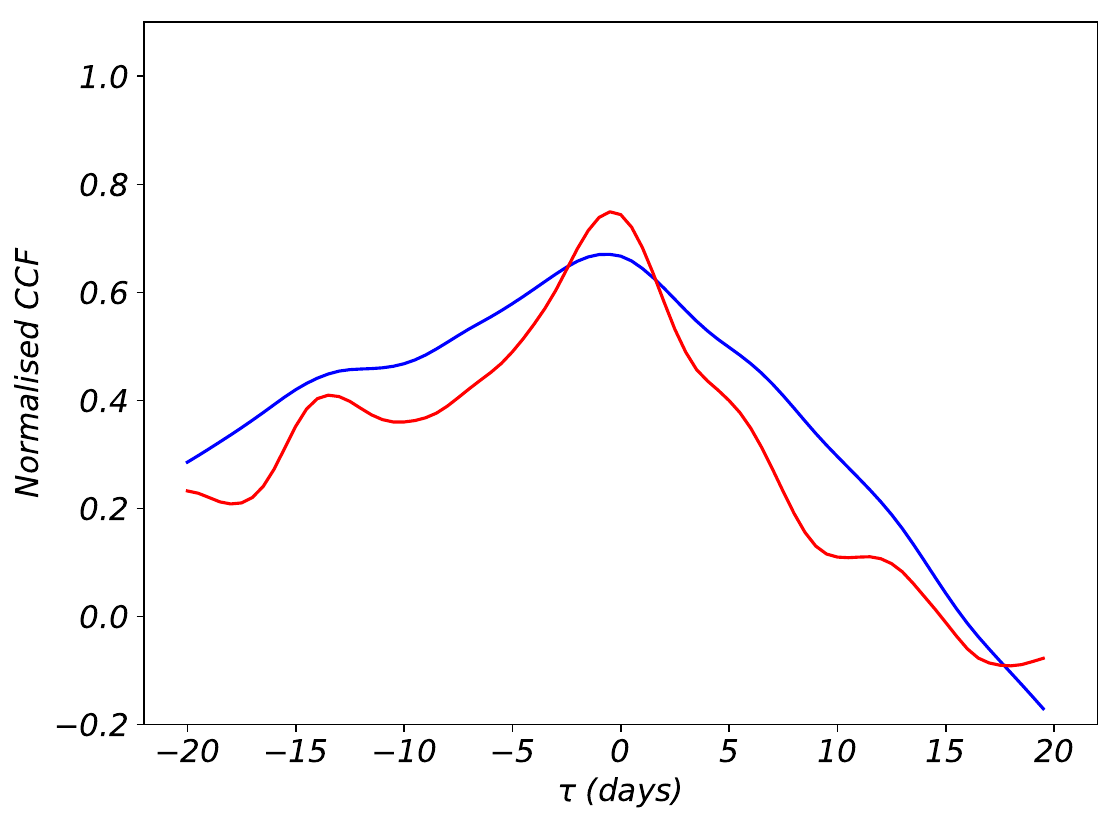}
	\caption{Top panel: Hard X-ray lightcurve from \textit{Swift}. Middle panel: Blue lightcurve is observed UVM2 flux. Red shows the simulated lightcurve from disc reprocessing of hard illuminating flux, with additional intrinsic soft variability. Bottom panel: CCFs for these lightcurves, blue for observed CCF, red for simulated CCF.}
	\label{fig:reprocessing_largevariablehard_flatsoft_flatdisc_LCs}
\end{figure}

\begin{figure}
	\includegraphics[width=\columnwidth]{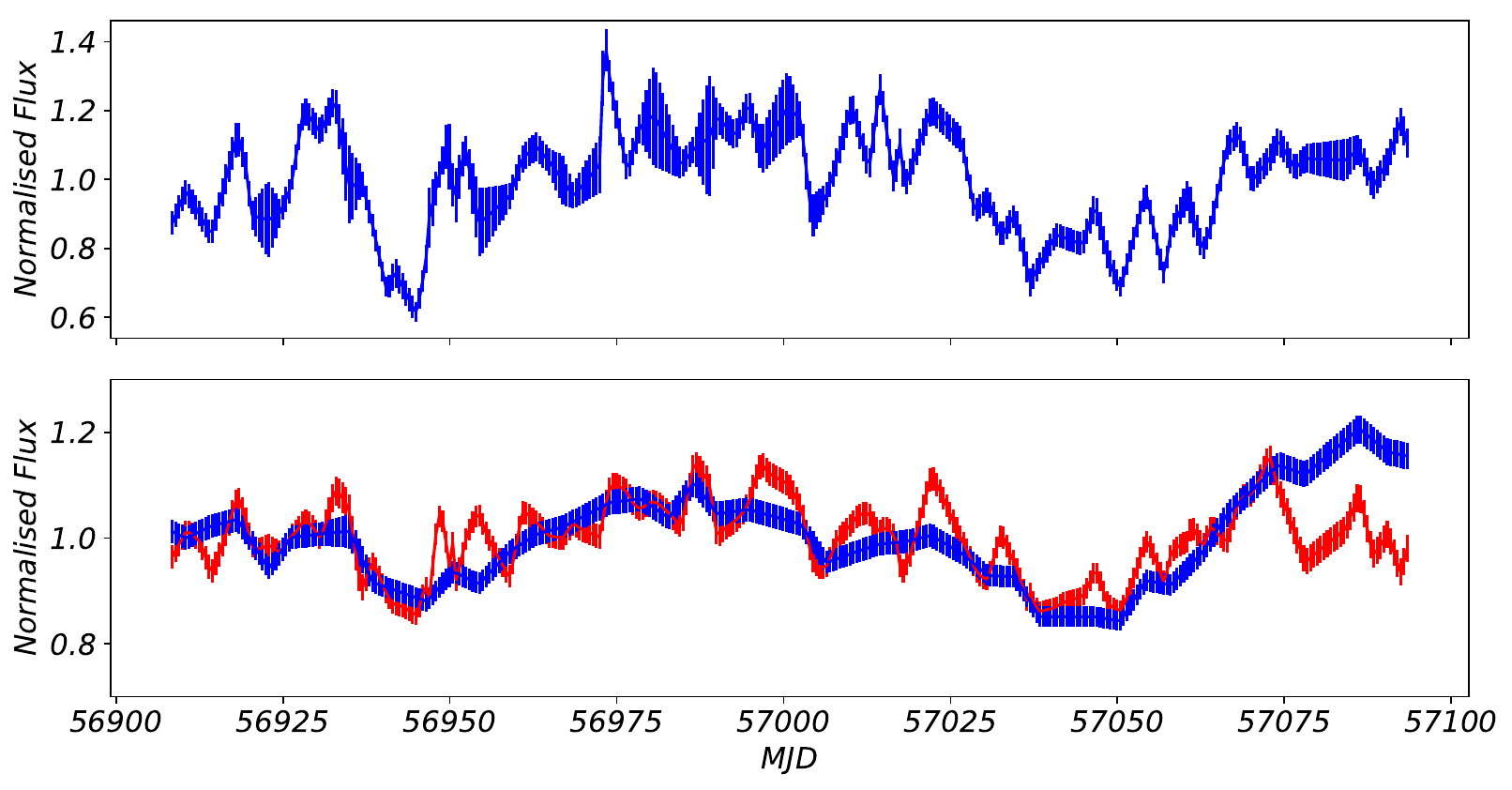}
	\includegraphics[width=\columnwidth]{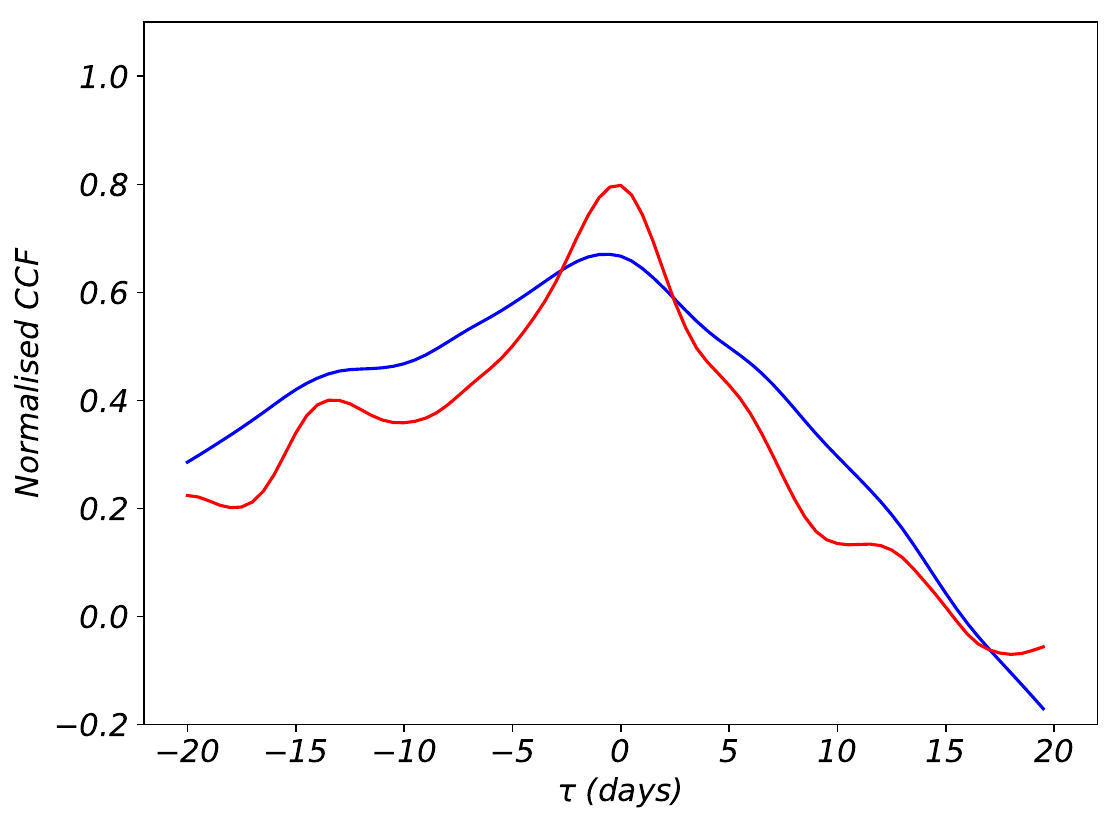}
	\caption{Results for \textbf{HX10+WC/2}. Top panel: Hard X-ray lightcurve from \textit{Swift}. Middle panel: Blue lightcurve is observed UVM2 flux. Red shows simulated lightcurve from disc reprocessing of hard illuminating flux, with intrinsic soft variations in the UV now suppressed (here by a factor $2$) to emulate component pivoting. Bottom panel: CCFs for these lightcurves, blue for observed CCF, red for simulated CCF.}
	\label{fig:reprocessing_largevariablehard_flatpivotingsoft_flatdisc_LCs}
\end{figure}

We now fold the additional variability - which is clearly present in the soft X-ray flux associated with the warm Compton component - into our model. The existence of this independent variability in the soft X-ray/UV component already challenges the above picture involving only reprocessing of hard fluctuations, as does simple energetics. The total flux in the low(high) state (including reprocessing from the source at height $10~R_g$) is 5.0(8.3)$\times 10^{-10}$~ergs~cm$^{-2}$~s$^{-1}$, with flux in the hot Compton of 1.2(2.0)$\times 10^{-10}$~ergs~cm$^{-2}$~s$^{-1}$. Thus the flux change in the hard X-rays is $0.8\times 10^{-10}$~ergs~cm$^{-2}$~s$^{-1}$, yet the change in the inferred optical/UV and soft X-ray components respectively are $0.6\times10^{-10}$~ergs~cm$^{-2}$~s$^{-1}$ and $1.9\times 10^{-10}$~ergs~cm$^{-2}$~s$^{-1}$. Clearly there must be intrinsic variability in the UV, as the flux change from hard X-rays alone is insufficient to drive these variations.

The accretion models used here associate the soft X-ray component with the high energy emission of an optically thick, warm Comptonisation region: a part of the accretion disc where thermalisation is incomplete. This same component extends down to the UV in this model, so the UV and soft X-rays are connected. We assume here that the entire warm Compton emission follows the variability of the soft X-ray component, i.e. that this emission varies with fixed electron temperature and spectral index, contributing soft X-ray variability directly into the UVM2 band. This intrinsic disc emission is not predicted by the thin disc equations, but neither is the incomplete thermalisation required to form this warm Compton spectrum. We assume both arise from changes in the disc structure triggered by either the radiation pressure instability and/or atomic opacities (\citealt{JB20}).

We also include the hard X-ray reprocessing as well, reverting back to the expected hard X-ray scale height of $h_x=10$, which predicts only low level UV flux variability from hard X-ray reprocessing (see Fig.~\ref{fig:model_predictions_CARMA_HX}). We do not include the reprocessing of the soft Compton source on the outer disc, as the flat-disc geometry means that the outer disc subtends a very small solid angle to the soft Compton section of the disc. We refer to this model as \textbf{HX10+WC}. The results of this are shown in Fig.~\ref{fig:reprocessing_largevariablehard_flatsoft_flatdisc_LCs}, where we find the modeled light curve to have an rms error with respect to the observed of $9\times 10^{-2}$. 

We see that this matches the amplitude of longer timescale variation of UVM2, but again predicts too much fast variability. This can be seen in the CCFs as well. The simulation now matches quite well to the overall amplitude and width of the observed CCF, but has a narrower core, showing that the $4$~day timescale characteristic of the warm Compton component is too prominent (see the ACF in Fig~\ref{fig:U_UVM2_SX_HX_COMPS_CCFs}). 

Thus if we assume that the entire component spanning soft X-ray to UV energies varies together as a single structure, changing only in normalisation - the intrinsic variability of the soft X-ray excess - this gives far too much fast variability to the resultant UVM2 curve, although it can match the amplitude and shape of the slower variability.

We can suppress some of the amplitude of variability by assuming the warm Comptonisation is not only changing just in normalisation but also in spectral index and/or electron temperature, so that the intrinsic soft variations in UVM2 are smaller than those in the soft X-ray band. We reduce the amplitude of input variations in UVM2 by an arbitrary factor $2$ to see the impact of this effect, yielding Fig.~\ref{fig:reprocessing_largevariablehard_flatpivotingsoft_flatdisc_LCs}. We term this model \textbf{HX10+WC/2}.

This is definitely the best match to the data so far, with an rms error with respect to the data of $6\times10^{-2}$. However there is still more fast UVM2 variability predicted than observed, as seen both by comparing the lightcurves and by the narrow, mildly correlated core of the predicted CCF ($CCF^{max}_{UVM2, HX}=0.80$. 

\section{Additional reprocessing from an inner BLR/wind}
\label{section:BLR_reprocessing}
The data are thus broadly consistent with the majority of the UVM2 variability being {\em intrinsic} to the warm Comptonisation region. However, the models where this warm Comptonisation region spans between UV and soft X-rays, and varies like the soft X-rays, results in too much variability in the UV on $\sim$day timescales, even considering moderate spectral pivoting. Thus, we now explore if part of the UVM2 variability can instead be explained by reprocessing on larger scale structures. Ark~120 clearly has a BLR which subtends a substanttial solid angle to the central source, so there should be a UV reprocessed component from this, contributing to the variability, in addition to that arising from an intrinsic disc origin. In NGC~4151 this was the only significant reprocessed contribution to the UV, with no evidence for a disc on smaller scales (MD20).

We first test a model where we assign half of the UVM2 flux to reprocessing in an extended region like the BLR or a wind on its inner edge (see Fig.~\ref{fig:geom}c). This means that the underlying accretion flow model changes, since half of the UVM2 flux is now not connected to direct emission from the disc. A minimal change to the mean {\tt agnsed} model which halves the UV flux while maintaining the X-ray flux can be produced by reducing $\log(L/L_{Edd})$ to $-1.45$ from the SED mean of $-1.31$ while increasing $R_{hot}$ from $23.1$ to $35~R_g$, i.e. mainly reducing the soft x-ray excess. 

\begin{figure}
    \centering
    \includegraphics[width=\columnwidth]{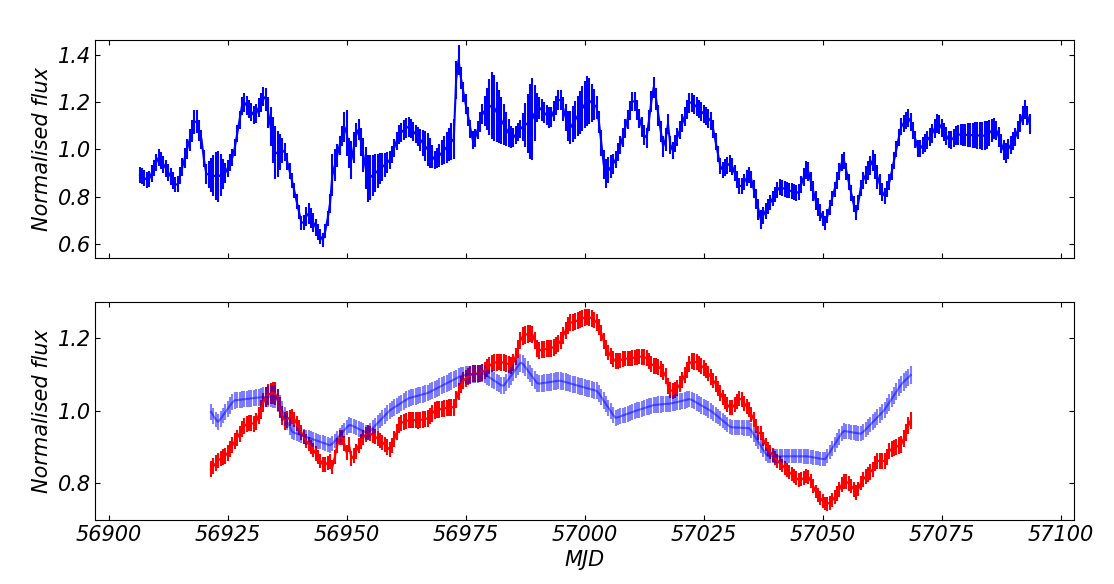}
    \includegraphics[width=\columnwidth]{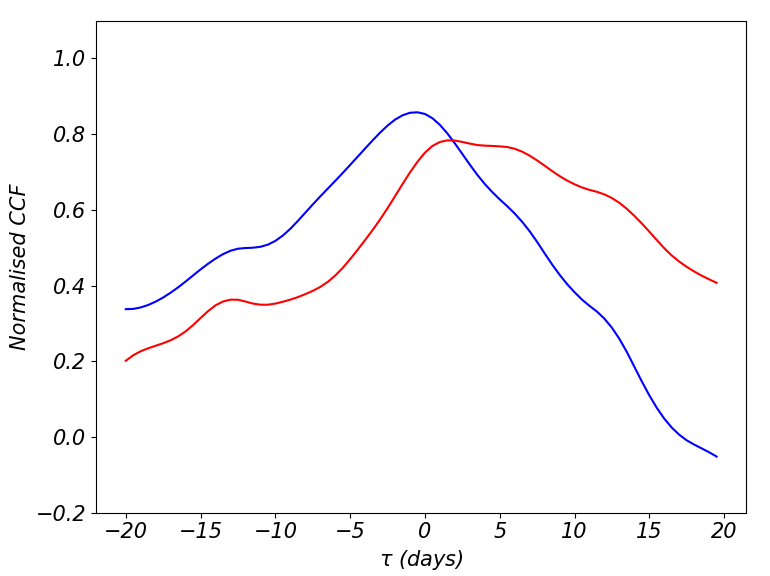}
    \includegraphics[width=\columnwidth]{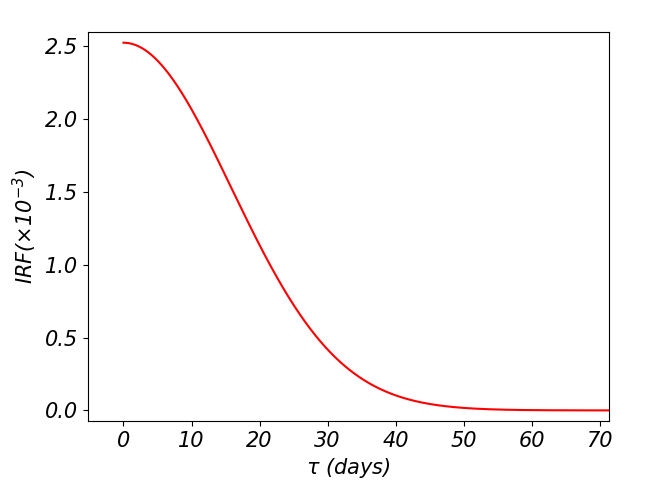}
    \caption{Results of model fit where $50~\%$ of the UVM2 flux goes like the results of \textbf{HX10+WC/2} (i.e. like Fig.~\ref{fig:reprocessing_largevariablehard_flatpivotingsoft_flatdisc_LCs}), while the other $50\%$ is produced by a fit impulse response function convolved with the luminosity-weighted sum of the soft and hard Compton component curves. The proportions of each Compton component curve fed into the $IRF$ convolution are selected such that they match the proportions expected from the SED. Top panel shows observed hard component light curve. Second panel (blue) shows observed UVM2 curve, while red shows our resultant simulated UVM2 light curve. Third panel shows the cross correlation functions with the usual observed (blue) and simulated (red) colour schemes. Fourth panel shows the best-fitting $IRF$ with which half of the driving X-ray flux is convolved.}
    \label{fig:reprocessing_largevariablehard_flatpivotingsoft_flatdisc_BLR_wSX_input}
\end{figure}

Fig.~\ref{fig:geom}c then implies that the wind will reprocess both the hard X-ray and the warm Comptonisation components. Section~\ref{sec:model2} has already shown the tension in the warm Comptonisation models for the soft X-ray excess, in that the observed soft X-ray flux shows much more fast variability than the observed UV flux. Models where the soft Compton varies only in normalisation clearly show too much fast variability (Fig.~\ref{fig:reprocessing_largevariablehard_flatsoft_flatdisc_LCs}). Even assuming that the warm Comptonisation changes in spectral index such that its contribution to the UV variability is only half as large as that in the soft X-rays shows moderate tension with the level of fast variability in the UV data (Fig.~\ref{fig:reprocessing_largevariablehard_flatpivotingsoft_flatdisc_LCs}). Here, since half of the UV lightcurve is assumed to be from reprocessing in a more extended region, then the UV variability of directly observed warm Comptonisation is diluted, so should give a better match to the data.

We use the hot Compton plus reduced variability warm Comptonisation lightcurve (\textbf{HX10+WC/2}) as our variable illumination. We assume that all of this is reprocessed in a large scale region, and give maximal freedom on this by convolving the lightcurve through an impulse response function ($IRF$). We assume this $IRF$ has a Gaussian shape, but truncate this for lags below zero to maintain causality. The UVM2 lightcurve is then modelled assuming that half of its emission coming from that $IRF$ convolution with the hard X-rays, and half from the simulated lightcurve of the pivoting warm Comptonisation model (i.e. the model UV lightcurve in Fig.~\ref{fig:reprocessing_largevariablehard_flatpivotingsoft_flatdisc_LCs}). The parameters describing the $IRF$ are then optimised by fitting this modelled UVM2 curve to the data.

Fig.~\ref{fig:reprocessing_largevariablehard_flatpivotingsoft_flatdisc_BLR_wSX_input} shows
the results for the best fit $IRF$. The driving hard X-ray lightcurve (blue) is in the upper panel, while the next panel shows the predicted UVM2 lightcurve (red) versus that observed (blue). The short term variability is mostly suppressed, but the predicted lightcurve does not match well to that observed, especially around MJD57000; the overall rms error is quite high, at $0.11$. The next panel shows this mismatch more clearly in terms of the CCF between the hard X-ray and UVM2 bands, with the predicted UVM2 (red) showing more lagged emission than the real data (blue). This highlights an issue with using the large scale reprocessor as the source of smoothing for the UVM2 lightcurve. Reprocessing on the light travel time inherently gives a light curve which is lagged on timescales similar to the smoothing timescale. This predicted lag can clearly be seen in the predicted CCF (red), but is not actually present in the observed data (blue). The best fit reconstructed $IRF$ (bottom) peaks at zero lag, but the mean lag is large in order to smooth the X-ray lightcurve to a level similar to that of the UVM2 data, but this imprints a lag which is not seen in the data. 

\begin{figure*}
    \centering
    \includegraphics[width=\textwidth]{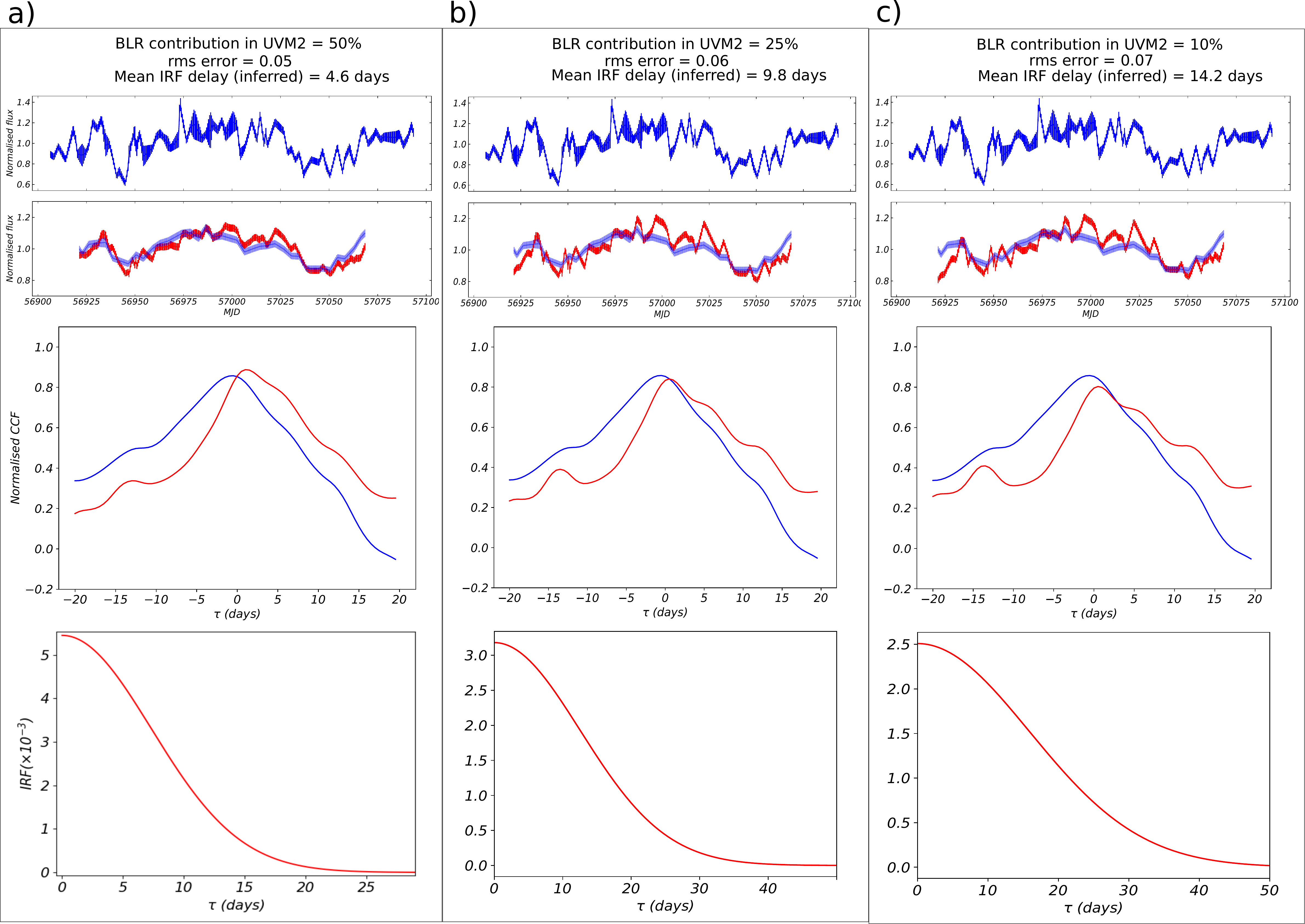}
    \caption{Results of model fits where a percentage of the UVM2 contributions is from the hard X-rays (only) being reprocessed through some new impulse response function ($IRF$), (which is allowed to vary in shape such that the simulated light curve best fits the data), while the remained of the UVM2 goes like the results of  \textbf{HX10+WC/2} (i.e. like Fig.~\ref{fig:reprocessing_largevariablehard_flatpivotingsoft_flatdisc_LCs}). Top rows show the observed hard component light curve. Second row panels show observed UVM2 curve in blue, and simulated UVM2 in red. Third row panels shows the cross correlation functions with the usual observed (blue) and simulated (red) colour schemes. Fourth row panels show the best-fitting $IRFs$ which result from the UVM2 curve fit.}
    \label{fig:reprocessing_largevariablehard_flatpivotingsoft_flatdisc_BLR}
\end{figure*}

In Fig.~\ref{fig:reprocessing_largevariablehard_flatpivotingsoft_flatdisc_BLR}a, we try the same approach, with $50\%$ of the UV variations produced as before by \textbf{HX10+WC/2} but where the $IRF$ now reprocesses only the hard X-ray Comptonisation. This gives the best match so far to the observed long term trends in the UV lightcurve (with an rms error of $5\times10^{-2}$). However the CCF still shows the same key discrepancy, in that all these models with a large scale reprocessor predict a lag on the same timescale as the smoothing, unlike the data. Further, when more complicated $IRF$ forms are tested in the same way as above (e.g. two or three Gaussians instead of one, cf. MD20), these results do not improve; the $IRF$ optimises to the same zero-peaked single Gaussian form to best match the UVM2 curve.

As well as testing a UVM2 light curve composed of $50\%$ BLR-reprocessed X-rays, we can also explore how the behaviour of the model changes when this proportion is reduced. In Fig.~\ref{fig:reprocessing_largevariablehard_flatpivotingsoft_flatdisc_BLR}b we show the model results for an IRF optimised to a BLR reprocessing contribution to the UVM2 of $25\%$, and in Fig.~\ref{fig:reprocessing_largevariablehard_flatpivotingsoft_flatdisc_BLR}c we show the case for only $10\%$. We see that, as the BLR contribution in the UVM2 decreases, we increase the direct component of the warm Comptonisation which has too much fast variability for the data.

Existing size estimates for the BLR layers which produce X-ray broad emission lines suggest a minimum radius of $20$~ld (\citealt{NPR16}, using the iron K$\alpha$ line FWHM from Chandra/HETG spectra). Meanwhile, H$\beta$ delays with respect to the continuum suggest a size scale for the optical BLR of $40$~ld (\citealt{PG91}). The BLR size scale from our model with only $10\%$ UVM2 BLR contribution is in closest agreement with these estimates. However those predictions are, themselves, model-dependent, with some inferring the size scale from the diffuse continuum from the BLR, while others have it from a wind inwards of the BLR.

\section{Discussion and Conclusions}
\label{sec:Conclusions}

Intensive continuum reverberation mapping campaigns, with simultaneous UV and X-ray data, are giving us a new way to probe the accretion flow geometry in AGN. The SED clearly changes with $L/L_{Edd}$, so we expect that the reverberation signals should also change. 

So far, most AGN monitored have been at low $L/L_{Edd}$. For example, MD20 used the NGC~4151 data to probe the accretion structure at $L/L_{Edd}\sim 0.015$. Here, the source SED peaks at $100$~keV, so the bolometric flux is dominated by the hard X-ray component, with a weaker peak in the UV. Thus, energetically, the hard X-ray reprocessed flux on the disc can produce a large UV variable flux, but this UV reprocessed flux varies too fast unless the disc is truncated below $400~R_g$. This used to be very controversial, as X-ray spectral fitting showed a broad residual even after accounting for the (marginally) resolved line seen in the Chandra data for NGC~4151 (e.g. \citealt{KBB15}). Most importantly, this appeared to vary as expected from extreme inner disc reverberation (\citealt{Zoghbi2012}). However, this is now seen as much more likely to be associated with complex and variable X-ray absorption, with no requirement for a relativistic reflection component from the inner disc (\citealt{ZMC19}). This material is likely to be the inner BLR and/or a wind on the inner edge of the BLR (see also \citealt{DFP19}), which likely also produces most of the reverberating UV in this source (MD20). 

This is in contrast to the results shown here for Ark~120. This AGN has $L/L_{Edd}\sim 0.05$, and its SED is dominated by a peak in the UV which is larger than the hard X-ray component, so the X-rays are not energetically dominant. There is clearly some intrinsic inner disc, probably extending down to $\sim 20-30R_g$ as is probably seen in the moderately broadened reflected spectrum (P18), consistent with the size scale of disc truncation derived in the {\sc agnsed} model in order for accretion flow to power the observed X-ray emission. We explore whether the variable UV can be produced by reprocessing of the X-rays in a range of potential geometries and scenarios. First we test whether reprocessing is from an X-ray source of size $\sim 10~R_g$, but this fails badly, both in under-predicting the amount of UV variability seen, and in reproducing far too much of the fast X-ray variability in the UV lightcurve (Fig.~\ref{fig:model_predictions_CARMA_HX}). Instead, we try a much larger scale height for the X-ray source (`lamppost'), at $\sim 100~R_g$ above the black hole. This gives a larger amplitude of variability from reprocessing, but there is still too much fast variability retained as the light travel times are too short ($\sim1$ day for a $100~R_g$ scale height and $30~R_g$ to the inner disc edge: Fig.~\ref{fig:reprocessing_variablehard_flatsoft_flatdisc_LCs_HX100_corrected}).

Then we allow some fraction of the warm Comptonisation to directly emit in the UVM2 bandpass. This already implies that some of the UV variability is intrinsic, rather than simply arising from reprocessing. This increases the amount of variability in the UV, but the warm Compton component typical variability timescales are only slightly longer than those of the hard Compton, so this produces far too much fast variability in the predicted UV flux (Fig 8). Assuming that the warm Comptonisation component variability comes from pivoting rather than a simple change in overall normalisation means that the amplitude of this direct component in the UV can be tuned to the data. We find a fairly good match if the UV changes by half the amplitude of the warm Compton component seen in the soft X-rays, but the models still appear to have too much fast variability (Fig 9). We allow some fraction of the UV variability to be produced by a large scale reprocessor connected to the BLR or an inner wind. We use a transfer function to give maximal flexibility to the geometry/spectrum of this reprocessor. We assume half of the UV flux arises from this large size scale, while half arises from the accretion flow, and calculate the UV variability arising from disc reprocessed hard X-rays (assuming $H_x=10R_g$) and the direct, but reduced amplitude variable emission from the warm Compton component. We assume the large scale reprocessor sees both hot corona and warm Compton. This somewhat overproduces the amount of variability, and also over-predicts the observed lags (Fig.~10). The UV lightcurves have a typical timescale of $\sim 10-20$ days, but smoothing of the observed hard X-ray lightcurve on these timescales also implies a similar lag timescale, which is not present in the data, which is a generic issue for a range of models with a large scale reprocessor (Fig.~11). 

We conclude that most of the UV variability must be intrinsic rather than reprocessed. A standard accretion disc cannot vary on these timescales (see e.g. the discussion in \citealt{ND18}), but the structures required to produce the observed emission do not look like a standard accretion disc but rather are better described by a warm Comptonisation component. However, in the models this spans between the soft X-ray and UV as a single component, but the soft X-ray excess variability is much faster than the observed UV variability. If this is a single component, then we require that the soft X-ray excess is the more variable high energy tail. 

Thus our (lack of) understanding of the UV variability is directly related to our (lack of) understanding of the warm Comptonisation component. This is not seen in the stellar mass black hole binary systems at the same $L/L_{Edd}$ (see e.g. \citealt{DGK07}), so pointers to the nature of this component may come from the difference between stellar and supermassive black holes. The main change with mass is the temperature of the optically thick accretion material. In the stellar mass systems this is $\sim 10^7$~K, peaking in the X-ray band, while for AGN it reduces to $\sim 10^5$~K, peaking in the UV. Atomic opacities are very important at UV temperatures, but have very little effect at X-ray temperatures. The UV opacity could induce convective turbulence in the disc, and/or power winds which fail and crash back down to the surface, shock heating it to produce all or part of the soft X-ray excess (\citealt{GD17}). 

Another difference is in the extent of the radiation pressure dominated section of the disc. This is not the same as the Eddington limit, where the radiation pressure exceeds gravity. Instead, there is an instability which occurs when the radiation pressure exceeds gas pressure inside the disc. The ansatz of a standard Shakura-Sunyaev disc is that the vertically integrated heating rate follows the total pressure, so is $\propto T$ in the gas pressure dominated regime, but $\propto T^4$ in the radiation pressure dominated region. A small fluctuation in temperature leads to a larger fluctuation in pressure in the radiation dominated regime, which leads to a runaway increase in heating. The ratio of radiation to gas pressure goes as $M^{1/4}$ in a Shakura-Sunyaev disc, so is approximately 100 times larger in AGN compared to a BHB at the same Eddington fraction. The standard disc models predict that the inner disc in BHB should also be subject to this instability at $L/L_{Edd}\ge 0.05$, yet the observed BHB discs are extremely stable (see e.g. \citealt{GD04}). This motivates different heating prescriptions, but it remains clear that AGN discs are much more strongly affected by the radiation pressure instability than BHB (see e.g. \citealt{GJCW17}).

\cite{JB20} include both the UV opacities and radiation magneto-hydrodynamics to capture the full heating rates from magnetic re-connection within the accretion flow. They find that there is some non-linear outcome of atomic opacities coupled to the radiation pressure instability which changes the entire vertical structure of the disc, and that this optically thick flow can vary on the thermal timescale rather than the much slower viscous timescales expected in the Shakura-Sunyaev disc.

We speculate that one or more of these processes is responsible for the intrinsic UV variability, on a timescale which is slow compared to the X-rays. The much faster variability of the soft X-ray excess could be produced by shocks from a failed UV line driven disc wind. The hard X-rays have their own even faster variability, but also see the UV and soft X-rays as seed photons, so the long term X-ray variability lags the UV on the light travel time, whilst also producing reprocessed UV variability which lags the X-rays on a similar timescale. 

Better monitoring data now exist for Fairall 9, an AGN with similar parameters and a similar SED (\citealt{HSEH20}). This will be used to explore these possibilities in subsequent work (Hagen et al., in prep.).

\section*{Acknowledgements}
RDM acknowledges the support of a Science and Technology Facilities Council (STFC) studentship through grant ST/N50404X/1. CD acknowledges the STFC through grant ST/T000244/1 for support, and multiple conversations about the Fairall 9 data with Scott Hagen, and Rick Edelson.

\section*{Data Availability}
X-ray and UV data underlying this article are available at NASA's HEASARC archive (https://heasarc.gsfc.nasa.gov/cgi-bin/W3Browse/w3browse.pl). Access to the reprocessing code is available on request from C.D. (chris.done@durham.ac.uk).


\bsp
\label{lastpage}
\end{document}